# A Reduced Cost Four-Component Relativistic Unitary Coupled Cluster Method for Molecules

Kamal Majee[1], Tamoghna Mukhopadhyay[1], Malaya K. Nayak[2,3,a] and Achintya Kumar Dutta[1,b]

[1]*Department of Chemistry, Indian Institute of Technology Bombay, Powai, Mumbai 400076, India*
[2]*Theoretical Chemistry Section, Bhabha Atomic Research Centre, Trombay, Mumbai 400085, India*
[3]*Homi Bhabha National Institute, BARC Training School Complex, Anushakti Nagar, Mumbai 400094, India*

## Abstract

We present a four-component relativistic unitary coupled cluster method for molecules. We have used commutator-based non-perturbative approximation using the "Bernoulli expansion" to derive an approximation to the relativistic unitary coupled cluster method. The performance of the full quadratic unitary coupled-cluster singles and doubles method (qUCCSD), as well as a perturbative approximation variant (UCC3), has been reported for both energies and properties. It can be seen that both methods give results comparable to those of the standard relativistic coupled cluster method. The qUCCSD method shows better agreement with experimental results due to better inclusion of the relaxation effects. A natural spinor-based scheme to reduce the computation cost of relativistic UCC3 and qUCCSD methods has been discussed.

a) Electronic mail: mknayak@barc.gov.in; mknayak@hbni.ac.in

b) Electronic mail: achintya@chem.iitb.ac.in

# 1. Introduction:

Accurate theoretical simulation of atoms and molecules containing heavy elements requires an electronic structure method that incorporates a balanced description of relativistic and electron correlation effects. The relativistic coupled cluster method[1,2] based on the four-component Dirac-Coulomb Hamiltonian[3,4] has emerged as one of the most promising candidates due to its systematically improvable nature. The relativistic coupled cluster method is generally used in singles and doubles approximation (CCSD). The extension to partial and full triples corrections[5–8], first and second-order property calculations[9–15], excited states[16–19], and transition property calculations[20–23] have been reported in the literature.

The standard relativistic coupled cluster method is non-hermitian and non-variational. One needs to solve an extra set of equations to make the coupled cluster energy functional stationary[24,25] with respect to the external perturbation. Alternative ansatzes like expectation value coupled cluster (XCC)[26] or extended coupled cluster (ECC)[27,28] have been used in the context of relativistic calculations[29,30]. However, none of these approaches are hermitian in nature. Using a unitary ansatz can lead to a hermitian formulation of the coupled cluster method[31,32], which can mitigate the problems associated with the standard non-hermitian formulation of a relativistic coupled cluster and allows one to describe the same-symmetry conical intersections using the coupled cluster method[33]. One of the major problems with the unitary coupled cluster (UCC)[32,34–36] is the non-terminating expansion of the corresponding similarity-transformed Hamiltonian. A forceful truncation would lead to a loss of size-extensivity of the energy[32]. Bartlett and co-workers[34] have proposed a truncation scheme based on the perturbative analysis of the UCC energy functional, which retains the size extensivity of the energy. Taube and Bartlett[35] put forward an alternative method of truncation that guarantees exact results for two-electron systems. However, the performance of the perturbative approximation to the UCC theory deteriorates for complex molecules due to the lack of smooth convergence of the underlying lower-order Moller Plesset (MP) perturbation theory[33,37]. One can circumvent this problem by using a non-perturbative commutator-rank truncation[36,38,39] schemes, where the similarity transformed Hamiltonian is truncated up to a particular commutator-rank. Sur *et. al.*[36] have reported a relativistic UCC method for atoms using the commutator-rank truncation schemes. However, applications to molecules have additional challenges. The higher number of correlated electrons and larger size of the basis set can significantly increase the computational cost of relativistic UCC calculation for molecules. The recently proposed natural spinor framework has been shown to reduce the computational cost of the standard relativistic coupled cluster method with systematically controllable accuracy[16,40,41]. The relativistic UCC method can also benefit from the use of frozen natural spinors. There has been growing interest in the unitary coupled cluster (UCC) theory in recent times due to its implication in the field of quantum computing[42–46]. However, an efficient relativistic UCC method, even on a classical computer has great potential for application. The aim of this manuscript is to describe the theory and implementation of a low-cost relativistic unitary coupled cluster method for molecules.

## 2. Theory and Computational details:

### 2.1 Relativistic Unitary Coupled Cluster Theory

The Dirac-Coulomb Hamiltonian[3] $\hat{H}_{DC}$ for a molecular system under the Born-Oppenheimer approximation can be defined as

$$\hat{H}_{DC} = \sum_{i}^{N}\left[c\vec{\alpha}_i\cdot\vec{p}_i + \beta_i m_0 c^2 + \sum_{A}^{N_{nuc}} V_{iA}\right] + \sum_{i<j}^{N} \frac{1}{r_{ij}} I_4. \tag{1}$$

Where, $\hat{p}_i$ and $m_0$ are the momentum operator and rest mass of the electron, respectively. Here, $c$ is the speed of light. $V_{iA}$ denotes the potential energy operator for the $i$-th electron in the field of nucleous $A$. $\alpha$ and $\beta$ are the Dirac matrices. $I_4$ is the $4\times 4$ identity matrix. The reference ground state wave function $\left(|\phi_0\rangle\right)$ for the many-electron system is obtained by using the mean-field approximation[3]. The Dirac-Hartree-Fock (DHF) equation can be written in the matrix form as

$$\begin{bmatrix} \hat{V} + \hat{J} - \hat{K} & c(\sigma_{psm}\cdot\hat{p}) - \hat{K} \\ c(\sigma_{psm}\cdot\hat{p}) - \hat{K} & \hat{V} - 2m_0 c^2 + \hat{J} - \hat{K} \end{bmatrix} \begin{bmatrix} \phi^L \\ \phi^S \end{bmatrix} = E \begin{bmatrix} \phi^L \\ \phi^S \end{bmatrix}. \tag{2}$$

In the above matrix form, the large and small components of the spinor $\phi$ are expressed by $\phi^L$ and $\phi^S$ respectively, where each component itself has a form of two-spinor. The nuclear-electron interaction is denoted by $\hat{V}$, whereas the Coulomb and the exchange operators are represented as $\hat{J}$ and $\hat{K}$, respectively and $\sigma_{psm}$ are the Pauli's spin matrices. The electron correlation effect is included using the UCC theory[33], where the exact ground state wave function $|\psi\rangle$, is obtained by applying an exponential unitary operator on the reference ground state wave function

$$|\psi\rangle = e^{\hat{\sigma}}|\phi_0\rangle \tag{3}$$

The $\hat{\sigma}$ is the anti-hermitian $\left(\hat{\sigma}^\dagger = -\hat{\sigma}\right)$ unitary coupled cluster operator and can be expanded up to the excitation manifold.

$$\hat{\sigma} = \hat{\sigma}_1 + \hat{\sigma}_2 + \hat{\sigma}_3 + ..... + \hat{\sigma}_N \tag{4}$$

Where,

$$\hat{\sigma}_n = \left(\frac{1}{n!}\right)^2 \sum_{ijk....abc....}^{n} \sigma_{ijk........}^{abc......} \left\{\hat{a}_a^\dagger \hat{a}_b^\dagger \hat{a}_c^\dagger .....\hat{a}_i \hat{a}_j \hat{a}_k\right\} = \hat{T}_n - \hat{T}_n^\dagger \tag{5}$$

Where $\hat{T}_n$ is the $n$-fold standard coupled cluster operator.

The ground state correlation energy in the four-component relativistic UCC method is obtained as

$$E_{Correlation} = \langle \phi_0 | \bar{H}_{DC}^{UCC} | \phi_0 \rangle \tag{6}$$

Where,

$$\bar{H}_{DC}^{UCC} = e^{-\hat{\sigma}} \hat{H}_{DC} e^{\hat{\sigma}} \tag{7}$$

represents the similarity transformed Hamiltonian obtained by the unitary transformation of the no-pair approximated[3,4] normal-ordered Dirac-Coulomb Hamiltonian $\left( \hat{H}_{DC} \right)$. The anti-hermitian nature of $\hat{\sigma}$ ensures the hermiticity of the $\bar{H}_{DC}^{UCC}$, which is in contrast to the non-hermitian nature of the standard relativistic coupled cluster similarity transformed Hamiltonian $\left( \bar{H}_{DC} \right)$

$$\bar{H}_{DC} = e^{-\hat{T}} \hat{H}_{DC} e^{\hat{T}} \tag{8}$$

The unitary coupled cluster operator $\hat{\sigma}$ is generally truncated up to singles and doubles excitation levels, and it leads to the UCCSD method,

$$\hat{\sigma} = \hat{\sigma}_1 + \hat{\sigma}_2 \tag{9}$$

Where,

$$\hat{\sigma}_1 = \sum_{ia} \sigma_i^a \{\hat{a}_a^\dagger \hat{a}_i\} - \sum_{ia} \left(\sigma_i^a\right)^* \{\hat{a}_i^\dagger \hat{a}_a\} \tag{10}$$

$$\hat{\sigma}_2 = \frac{1}{4}\left[ \sum_{ijab} \sigma_{ij}^{ab} \{\hat{a}_a^\dagger \hat{a}_b^\dagger \hat{a}_j \hat{a}_i\} - \sum_{ijab} \left(\sigma_{ij}^{ab}\right)^* \{\hat{a}_a \hat{a}_b \hat{a}_j^\dagger \hat{a}_i^\dagger\} \right] \tag{11}$$

Where $i, j, k, l$ and $a, b, c, d$ symbols are used to denote the occupied and virtual spinors, respectively. $\sigma_i^a$ and $\sigma_{ij}^{ab}$ are the singles and doubles cluster amplitudes.

The singles and doubles cluster amplitudes are calculated by left projecting the $\bar{H}_{DC}^{UCC}$ with excited state determinants as,

$$\langle \phi_i^a | \bar{H}_{DC}^{UCC} | \phi_0 \rangle = 0 \tag{12}$$

$$\langle \phi_{ij}^{ab} | \bar{H}_{DC}^{UCC} | \phi_0 \rangle = 0 \tag{13}$$

Where $|\phi_i^a\rangle$ and $|\phi_{ij}^{ab}\rangle$ are the singly and doubly excited determinants with respect to the DHF configuration, respectively.

The relativistic UCC similarity transformed Hamiltonian can be expanded using the Baker-Campbell-Hausdorff (BCH) formula as,

$$\bar{H}_{DC}^{UCC} = \hat{H}_{DC}^{UCC} + \left[\hat{H}_{DC}^{UCC}, \hat{\sigma}\right] + \left[\left[\hat{H}_{DC}^{UCC}, \hat{\sigma}\right], \hat{\sigma}\right] + \left[\left[\left[\hat{H}_{DC}^{UCC}, \hat{\sigma}\right], \hat{\sigma}\right], \hat{\sigma}\right] + \ldots \quad (14)$$

Unlike the traditional CC theory[32], the BCH expansion doesn't have a natural truncation due to the presence of both excitation and de-excitation cluster operators.

One can derive commutator-based non-perturbative approximation using the "Bernoulli expansion"[38,47] of $\bar{H}_{DC}^{UCC}$. Assuming $|\phi_0\rangle$ to be the Dirac-Fock wave function, one can partition the no-pair approximated Hamiltonian into the Fock operator $\hat{F}$ and a fluctuation potential $\hat{V}$

$$\hat{H}_{DC} = \hat{F} + \hat{V} \quad (15)$$

The Fock operator in the four-component relativistic picture for a closed shell atom or molecule framework is block diagonal and rank-conserving operator[4].

$$\hat{F} = \sum_{ij} f_{ij} \{\hat{a}_i^\dagger \hat{a}_j\} + \sum_{ab} f_{ab} \{\hat{a}_a^\dagger \hat{a}_b\} \quad (16)$$

The fluctuation potential $\hat{V}$ for a Dirac-Coulomb Hamiltonian can be represented as,

$$\hat{V} = \frac{1}{4} \langle pq||rs\rangle \{\hat{a}_p^\dagger \hat{a}_q^\dagger \hat{a}_s \hat{a}_r\} \quad (17)$$

The $\hat{V}$ can be further partitioned into the non-diagonal $(\hat{V}_N)$ and the 'rest' part $(\hat{V}_R)$.

Following the reference[33,38], the $\bar{H}_{DC}$ can be expanded by the Bernoulli numbers

$$\bar{H}_{DC} = \bar{H}_{DC}^0 + \bar{H}_{DC}^1 + \bar{H}_{DC}^2 + \bar{H}_{DC}^3 + \ldots \quad (18)$$

$$\begin{aligned}
\bar{H}_{DC}^0 &= F + V, \\
\bar{H}_{DC}^1 &= [F, \hat{\sigma}] + \frac{1}{2}[V, \hat{\sigma}] + \frac{1}{2}[V_R, \hat{\sigma}], \\
\bar{H}_{DC}^2 &= \frac{1}{12}[[V_N, \hat{\sigma}], \hat{\sigma}] + \frac{1}{4}[[V, \hat{\sigma}]_R, \hat{\sigma}] + \frac{1}{4}[[V_R, \hat{\sigma}]_R, \hat{\sigma}], \\
\bar{H}_{DC}^3 &= \frac{1}{24}[[[V_N, \hat{\sigma}], \hat{\sigma}]_R, \hat{\sigma}] + \frac{1}{8}[[[V_R, \hat{\sigma}]_R, \hat{\sigma}]_R, \hat{\sigma}] + \frac{1}{8}[[[V, \hat{\sigma}]_R, \hat{\sigma}]_R, \hat{\sigma}] \\
&\quad - \frac{1}{24}[[[V, \hat{\sigma}]_R, \hat{\sigma}], \hat{\sigma}] - \frac{1}{24}[[[V_R, \hat{\sigma}]_R, \hat{\sigma}], \hat{\sigma}]
\end{aligned} \quad (19)$$

In the above equation, "$N$" encompasses all aspects related to the excitation and de-excitation parts of the target operator. On the other hand, "$R$" refers to the remaining segments of the operator, excluding the "$N$" components. The above expansion eliminates higher-than-linear commutators to the Fock operator and provides a concise basis for constructing approximations within the relativistic UCC method[33].

One can develop a non-perturbative approximation to UCC by taking the commutator up to a particular rank. For example, taking terms up to $\bar{H}_{DC}^3$ in the energy and $\bar{H}_{DC}^2$ in the amplitude equation leads to qUCCSD approximation [33].

$$E_{Gr}^{qUCCSD} = E^{DHF} + \langle \phi_0 | \bar{H}_{DC}^1 | \phi_0 \rangle + \langle \phi_0 | \bar{H}_{DC}^2 | \phi_0 \rangle + \langle \phi_0 | \bar{H}_{DC}^3 | \phi_0 \rangle \quad (20)$$

$$\langle \phi_i^a | \bar{H}_{DC}^1 + \bar{H}_{DC}^2 | \phi_0 \rangle = 0 \quad (21)$$

$$\langle \phi_{ij}^{ab} | \bar{H}_{DC}^0 + \bar{H}_{DC}^1 + \bar{H}_{DC}^2 | \phi_0 \rangle = 0 \quad (22)$$

The relativistic UCC3 method of Sur *et.al.* for atoms can be considered as a perturbative approximation to the qUCCSD method, where the amplitude and energy equations are restricted up to the third order and fourth order in perturbation, respectively. The explicit programmable expressions for four-component relativistic qUCCSD and UCC3 methods are provided in the supporting information.

Due to the presence of quadratic $\sigma_1^2$ terms, the qUCCSD method can be considered to be more complete than the UCC3 method and expected to give superior performance than the corresponding UCC3 method, especially when the starting Dirac-Fock determinant does not provide an appropriate zeroth-order description of the ground state wavefunction. The computational time in the qUCCSD and UCC3 method is dominated by $O(N_o^2 N_v^4)$ and $O(N_o^3 N_v^3)$ scaling terms, where $N_o$ and $N_v$ are the number of occupied and virtual spinors. The computation cost of four-component qUCCSD calculations for a closed shell molecule is at least 32 times higher than the corresponding non-relativistic calculation when Kramers symmetry is considered. However, in the present implementation, we have not considered the Kramers symmetry, as we want to keep the UCC3/qUCCSD codes modular and independent of how the underlying DHF spinors are generated. Therefore, the computational cost of the present implementation of the four-component UCC3 and qUCCSD is at least 256 times higher than the corresponding spin-summed non-relativistic version for closed shell molecules[4].

It is important to compare the computational cost of the relativistic UCC3 and qUCCSD methods with the standard relativistic coupled cluster method. Each iteration of standard CCSD equations contains one "ladder type" contraction containing four external integrals contributing to the doubles part of residual ($S_{ij}^{ab}$), which scales as $O(N_o^2 N_v^4)$

$$\frac{1}{2} \sum_{cd} \langle ab || cd \rangle \tau_{ij}^{cd} \rightarrow s_{ij}^{ab} \quad (23)$$

where the intermediate $\tau_{ij}^{ab}$ is defined as

$$\tau_{ij}^{ab} = t_{ij}^{ab} + t_i^a t_j^b - t_i^b t_j^a \tag{24}$$

In addition, there are two "ring type" contractions, which scale as $O(N_o^3 N_v^3)$

$$\sum_{me} t_{im}^{ae} W_{mbej} \to s_{ij}^{ab} \tag{25}$$

$$-\sum_{nf} \frac{1}{2} \left( t_{jn}^{fb} + t_j^f t_n^b \right) \langle mn || ef \rangle \to W_{mbej} \tag{26}$$

The above-mentioned ladder and ring type terms represent the most time-consuming steps in the coupled cluster calculations.

The qUCCSD method has two terms that contain four external integrals.

$$\frac{1}{2} \sum_{cd} \langle ab || cd \rangle \tau_{ij}^{cd} \to s_{ij}^{ab} \tag{27}$$

$$\frac{1}{4} \sum_{jbcd} \left( \sigma_j^c \right)^* \langle ac || bd \rangle \sigma_{ij}^{bd} \to s_i^a \tag{28}$$

Both terms in the present implementation scale as $O(N_o^2 N_v^4)$. In fact, the equation (28) can be calculated in a $O(N_o^2 N_v^3)$ step by first contracting the singles amplitude with the integral. However, such an approach will involve expanding the packed indices in four external integrals, resulting in enhanced I/O. The qUCCSD equations also contain the following nine " ring type" contributions,

$$-\sum_{jklbc} \left( \sigma_{jk}^{bc} \right)^* \langle bl || ji \rangle \sigma_{kl}^{ca} + \sum_{jkbcd} \left( \sigma_{jk}^{bc} \right)^* \langle ab || dj \rangle \sigma_{ki}^{cd} \to s_i^a \tag{29}$$

$$P(ij) P(ab) \sum_{kc} \langle ak || ic \rangle \sigma_{jk}^{bc} \to s_{ij}^{ab} \tag{30}$$

$$P(ij) P(ab) \frac{1}{3} \sum_{klcd} \langle kl || cd \rangle \sigma_{ik}^{ac} \sigma_{jl}^{bd} + P(ij) P(ab) \frac{1}{3} \sum_{klcd} \left( \sigma_{kl}^{cd} \right)^* \langle ad || il \rangle \sigma_{jk}^{bc} \to s_{ij}^{ab} \tag{31}$$

$$-P(ij) P(ab) \sum_{lcd} \left( \sigma_l^c \right)^* \langle ac || dj \rangle \sigma_{il}^{bd} + P(ij) P(ab) \sum_{klc} \left( \sigma_l^c \right)^* \langle bk || li \rangle \sigma_{jk}^{ca} \to s_{ij}^{ab} \tag{32}$$

$$-P(ij) P(ab) \sum_{klc} \langle kl || cj \rangle \sigma_l^b \sigma_{ik}^{ac} + P(ij) P(ab) \sum_{kcd} \langle kb || cd \rangle \sigma_j^d \sigma_{ik}^{ac} \to s_{ij}^{ab} \tag{33}$$

The most of the above terms scale as $O(N_o^3 N_v^3)$ except the first terms of equations (29) and (33) which alternatively can be implemented with $O(N_o^4 N_v^2)$ scaling. Therefore, the relativistic qUCCSD is approximately twice as costly as that of the standard relativistic CCSD method, which restricts the use of the relativistic qUCCSD method beyond small basis sets. Equations (28), (32), and (33) will not be present in relativistic UCC3, and its computational cost will be slightly smaller than standard relativistic CCSD because of the presence of fewer terms.

**2.2 Frozen Natural Spinors (FNS):**

Natural spinors are obtained by diagonalizing spin-coupled correlated one-particle reduced density matrix calculated using a relativistic electron correlation method[40]. The natural spinors are relativistic analogues of the natural orbitals proposed by Löwdin[48]. We have used the frozen natural spinors (FNS) framework in this work[40], where the occupied orbitals are kept frozen at their corresponding Dirac-Fock description, and the virtual orbitals are expanded in terms of natural spinors.

Virtual natural spinors are generated by the diagonalization of the virtual-virtual block of the correlated one-body reduced density matrix calculated at the relativistic MP2 level of theory.

$$D_{ab} = \sum_{cij} \frac{\langle ac||ij\rangle \langle ij||bc\rangle}{\varepsilon_{ij}^{ac} \varepsilon_{ij}^{bc}} \tag{34}$$

Where,

$$\varepsilon_{ij}^{ac} = \varepsilon_i + \varepsilon_j - \varepsilon_a - \varepsilon_c, \tag{35}$$

$$\varepsilon_{ij}^{bc} = \varepsilon_i + \varepsilon_j - \varepsilon_b - \varepsilon_c, \tag{36}$$

This $\varepsilon_i, \varepsilon_j, \varepsilon_a, \varepsilon_b$ and $\varepsilon_c$ are the DHF molecular spinor energies, whereas $\langle ac||ij\rangle$ and $\langle ij||bc\rangle$ are the antisymmetrized two-electron integrals.

One can obtain virtual natural spinors $V$ and their corresponding occupancies $n$ by diagonalizing the virtual-virtual block of the correlated one-body reduced density matrix,

$$D_{ab}V = Vn \tag{37}$$

These virtual natural spinors are sorted according to their occupancies, and virtual natural spinors above a specific cutoff are dropped. The rest of the natural spinors form $\tilde{V}$, where the tilde $(\sim)$ represents the truncated natural spinor basis.

The virtual-virtual block of the Fock matrix $(F_{VV})$ is then transformed to the truncated virtual natural spinor basis.

$$\tilde{F}_{VV} = \tilde{V}^{\dagger} F_{VV} \tilde{V} \tag{38}$$

The semi-canonical virtual natural spinors $(\tilde{Z})$ and their corresponding energies $(\tilde{\varepsilon})$ are obtained by diagonalizing the Fock matrix in the virtual natural spinor basis,

$$\tilde{F}_{VV} \tilde{Z} = \tilde{Z} \tilde{\varepsilon} \tag{39}$$

The transformation matrix ($B$) transforms the canonical DHF virtual spinor space to the semi-canonical natural virtual spinor space

$$B = \tilde{Z}\tilde{V} \qquad (40)$$

Therefore, our basis set is composed of the canonical DHF occupied spinors and semi-canonical natural virtual spinors generated by the frozen natural spinor (FNS) approximation.

The lower scaling four-component relativistic UCC3 and qUCCSD methods are implemented in the development version of our in-house quantum chemistry software package BAGH[49]. The BAGH is written primarily in Python, while Cython and Fortran have been used to write the most computationally demanding parts. BAGH depends upon other softwares for the integrals and converged SCF coefficients. The one and two-electron integrals and the converged four-component DHF spinors are taken from the PySCF[50]. In the first step, the partial integral transformation is performed to generate the integrals with two external indices in the molecular spinor basis.

$$\begin{aligned}(ia \mid jb) &= \sum_{\mu\nu\kappa\lambda} C_{i\mu}^{L*} C_{av}^{L} C_{j\kappa}^{L*} C_{b\lambda}^{L} \left(\mu^L \nu^L \mid \kappa^L \lambda^L\right) \\ &+ \sum_{\mu\nu\kappa\lambda} C_{i\mu}^{S*} C_{av}^{S} C_{j\kappa}^{S*} C_{b\lambda}^{S} \left(\mu^S \nu^S \mid \kappa^S \lambda^S\right) \\ &+ \sum_{\mu\nu\kappa\lambda} C_{i\mu}^{L*} C_{av}^{L} C_{j\kappa}^{S*} C_{b\lambda}^{S} \left(\mu^L \nu^L \mid \kappa^S \lambda^S\right) \\ &+ \sum_{\mu\nu\kappa\lambda} C_{i\mu}^{S*} C_{av}^{S} C_{j\kappa}^{L*} C_{b\lambda}^{L} \left(\mu^S \nu^S \mid \kappa^L \lambda^L\right)\end{aligned} \qquad (41)$$

Where molecular spinors are denoted by the label $i, j, a, b$ and $\mu, \nu, \kappa, \lambda$ denote the atomic spinors. The large and small basis components are represented by the superscript $L$ and $S$, respectively, and $C$ denote the transformation coefficient matrix of the atomic spinor to the molecular spinor basis. In the next step, the two external integrals are converted from Mulliken to Dirac notation and antisymmetrized.

$$\langle ij \| ab \rangle = (ia \mid jb) - (ib \mid ja) \qquad (42)$$

The MP2 energy, one-particle reduced density, and frozen natural spinors are calculated using equations $(34)$ to $(40)$. The transformation matrix from atomic spinor to the frozen natural spinor basis can be obtained as

$$C'_{p'\mu} = B_{p'x} C_{x\mu} \qquad (43)$$

In the next step, 0 to 4 external integrals are generated on the natural spinor basis as,

$$(p'r'|q's') = \sum_{\mu\nu\kappa\lambda} C_{p'\mu}^{'L*} C_{r'\nu}^{'L} C_{q'\kappa}^{'L*} C_{s'\lambda}^{'L} \left(\mu^L \nu^L | \kappa^L \lambda^L\right)$$

$$+ \sum_{\mu\nu\kappa\lambda} C_{p'\mu}^{'S*} C_{r'\nu}^{'S} C_{q'\kappa}^{'S*} C_{s'\lambda}^{'S} \left(\mu^S \nu^S | \kappa^S \lambda^S\right)$$

$$+ \sum_{\mu\nu\kappa\lambda} C_{p'\mu}^{'L*} C_{r'\nu}^{'L} C_{q'\kappa}^{'S*} C_{s'\lambda}^{'S} \left(\mu^L \nu^L | \kappa^S \lambda^S\right) \quad (44)$$

$$+ \sum_{\mu\nu\kappa\lambda} C_{p'\mu}^{'S*} C_{r'\nu}^{'S} C_{q'\kappa}^{'L*} C_{s'\lambda}^{'L} \left(\mu^S \nu^S | \kappa^L \lambda^L\right),$$

$$\langle pq \| rs \rangle = (pr|qs) - (pr|sq). \quad (45)$$

Subsequently, UCC3 and qUCCSD calculations are performed on the truncated FNS basis. Figure 1 gives a schematic description of FNS-based UCC3 and qUCCSD methods.

The most time-consuming steps in both four-component relativistic UCC3 and qUCCSD methods scale as $O(N_O^2 N_V^4)$ and $O(N_O^3 N_V^3)$. The effective number of virtuals in the frozen natural spinor basis is $N_{FNS}$ after dropping off $N_D$ virtual spinors

$$N_V = N_{FNS} + N_D \quad (46)$$

Therefore, the most time-consuming parts of our FNS-UCC3 and FNS-qUCCSD methods scale as $O\left(N_O^2 (N_V - N_D)^4\right)$ and $O\left(N_O^3 (N_V - N_D)^3\right)$.

**2.3 Perturbative Correction :**

A second-order perturbative correction for the virtual natural spinor truncation is added to the correlation energy obtained from UCC3 and qUCCSD methods. The MP2 correlation energy is calculated both in the untruncated canonical molecular spinor and truncated semi-canonical natural spinor basis. The difference between the two is taken as the correction and is approximated to be equivalent to the error caused by the truncation of virtual natural spinors.

$$\Delta E_{Corrected}^{UCC3/qUCCSD} = \Delta E_{Uncorrected(FNS)}^{UCC3/qUCCSD} + \left(\Delta E_{Canonical}^{MP2} - \Delta E_{FNS}^{MP2}\right) \quad (47)$$

Although we have restricted our attention to molecules in the present manuscript, our relativistic unitary coupled cluster codes and their FNS versions are general and equally applicable to atoms.

## 3. Results & Discussion:

### 3.1 Correlation energy:

As we will use FNS to reduce the computational cost of UCC3 and qUCCSD methods, it is essential to find an appropriate threshold that can balance computational cost and accuracy. To explore the convergence of the correlation energy with respect to the size of the virtual space in the FNS basis, we have plotted the correlation energy in relativistic UCC3 and qUCCSD methods with respect to the size of the virtual space for the HCl molecule. The uncontracted version of the aug-cc-pVTZ basis set is used for the calculation. Figure 2 shows that the correlation energy in both UCC3 and qUCCSD methods converge more quickly with respect to the size of the virtual space in the FNS basis than that observed in the canonical basis. It can be seen that the UCC3 correlation energy converges with 50% of the virtual spinors included in the calculation when the FNS basis is used. To recover the same amount of correlation energy, one must include at least 80% of the virtual spinors in the canonical basis. The use of MP2 perturbation correction improves the convergence behavior of FNS-UCC3 correlation energy. The correlation energy converges even with 20% of the total virtual space when MP2 correction is considered. The FNS-qUCCSD method also shows similar convergence behavior. The percentage of the virtual spinor is not an optimal criterion for the choice of the size of the virtual space in the natural spinor basis[40]. Therefore, we have chosen the FNS occupation number as a criterion for the selection of the size of the virtual space. Figure 3(a) presents the convergence of the correlation energy in FNS-UCC3 and FNS-qUCCSD methods with respect to the FNS occupation threshold. It can be seen that both FNS-UCC3 and FNS-qUCCSD correlation energy converge with the FNS occupation threshold of $10^{-6}$. The inclusion of perturbative correction further improves the convergence behavior of FNS-UCC3 and FNS-qUCCSD correlation energy, and they converge with an FNS threshold of $10^{-5}$ when MP2 perturbative correction is included. The frozen core (fc) approximation has been shown to have a prominent effect on the convergence behavior of the energy with respect to natural spinors. Figure 3(b) presents the convergence behavior of the correlation energy in frozen core approximation with respect to the FNS threshold. It can be seen that the correlation energy converges more quickly in both fc-FNS-UCC3 and fc-FNS-qUCCSD with respect to the FNS threshold, and a threshold of $10^{-5}$ is sufficient to get converged energy even without MP2 correction. The trend is consistent with that observed for the standard relativistic coupled cluster method[40]. The core level occupied spinors in a DHF determinant is related to the highest-lying unoccupied virtual spinors[40]. Consequently, freezing of the core level spinor reduces the importance of the corresponding virtual natural spinor in the correlation energy, and one can discard them without any significant loss of accuracy. This leads to quicker convergence of correlation energy with respect to the FNS threshold when frozen core approximation is used. Therefore, for the rest of the calculations, we have used the frozen core approximation unless it is explicitly specified.

To understand the suitability of the chosen occupation threshold, we have calculated the percentage of error in correlation energy recovery for the HX (X=F, Cl, Br, I) series of molecules in FNS-UCC3 and FNS-qUCCSD methods (See Table 1). An FNS occupation threshold of $10^{-5}$ is used for the calculations. Visscher and co-workers used the same set of molecules to understand the comparative effect of relativity and electron correlation on molecular properties[9]. The uncontracted aug-cc-pVTZ basis set (dyall.acv3z for Br and I) is used for the calculations. It can be seen that the occupation threshold $10^{-5}$ leads to recovery of a similar percentage (not identical) of the correlation energy for UCC3 and qUCCSD method. The percentage of recovered correlation energy is different for different molecules, but more than 99% of correlation energy is recovered for all four molecules. The correlation energy can be as high as 99.8 % in the case of HCl. Meanwhile, the percentage of correlation energy recovered is 99.3% for the HI. The number of active virtual spinors selected at the threshold of $10^{-5}$ for HI is 174, which is only 40% of the original virtual space of 434 spinors. The number of active virtual spinors in HBr is higher at 178, although the total number of virtual spinors is 362, which is less than that in HI. It can be seen that the error in the correlation energy recovered at the $10^{-5}$ threshold is less than 0.2 percent in both FNS-UCC3 and FNS-qUCCSD methods for all the molecules when MP2 correction has been included. Therefore, the FNS threshold of $10^{-5}$ can be considered as default when MP2 perturbative correction is incorporated. It should be noted that the convergence behavior of the correlation energy in the FNS-UCC3 and FNS-qUCCSD methods follows the same trend as that of the standard FNS-CCSD[40].

## 3.2 Dipole moment:

One of the advantages of the unitary coupled cluster is the ease of property calculation using the expectation value approach[36]. In this work, we have used a finite field approach for property calculations as it is more straightforward than the expectation value approach and suitable for quick benchmarking. The dipole moment values presented in this manuscript are calculated by adding a positive and negative electric field of 0.001 a.u. to the Hamiltonian. Table 2 presents canonical fc-UCC3 and fc-qUCCSD dipole moments for HX (X=F, Cl, Br, I). Uncontracted aug-cc-pVTZ basis set (dyall.acv3z basis set for Br and I ) has been used for the calculation. It can be seen that the canonical UCC3 and qUCCSD values are in good agreement with the standard CCSD values. The agreement is slightly better in the case of the qUCCSD method. Figure 4 presents the convergence of error in the dipole moment with respect to the percentage of virtual space in the canonical and FNS-based UCC3 and qUCCSD methods for HCl in the uncontracted aug-cc-pVTZ basis set. It can be seen that the dipole moment converges more slowly in the FNS basis than that observed in the canonical basis. Both UCC3 and qUCCSD dipole moments converge at around 70% of the virtual space in the FNS basis. The inclusion of the perturbative correction significantly improves the convergence in the FNS basis, and the error almost goes to zero with 40% of the virtual space. The poor performance in the FNS basis for the dipole moment is due to the fact that the ground state correlation energy and dipole moment behave very differently with respect to the basis set truncation[51]. The diffuse functions make a minimal contribution to the ground state correlation energy of neutral systems[52] and are generally among the first to be discarded by the FNS-based truncation scheme. On the other hand, these diffuse functions play a fundamental role in determining the accuracy of the dipole moment[53], and their truncation due to the FNS threshold leads to significant errors in the dipole moment. The trend is consistent with that observed in the standard FNS-CCSD[40] truncation[51]. Figure 5 presents the convergence of the UCC3 and qUCCSD dipole moments with respect to the occupation threshold using the frozen core approximation. It can be seen that both FNS-UCC3 and FNS-qUCCSD dipole moment converge with an occupation threshold of $10^{-5}$ when perturbative correction is included. The convergence is slower without the perturbative correction, and the dipole moment converges at a threshold of $10^{-6}$ for both FNS-UCC3 and FNS-qUCCSD. Table 2 shows that both the FNS-UCC3 and FNS-qUCCSD methods agree excellently with the corresponding canonical values, giving errors less than 0.002 Debye when perturbative corrections are included. In the absence of perturbative correction, the maximum error can be as high as 0.02 Debye. Both FNS-UCC3 and FNS-qUCCSD agree excellently with the experimental results for HF and HCl. However, their performance is slightly inferior for HI, where the error with respect to the experiment is 0.04 Debye. The CCSD method gives almost identical performance as that of the corresponding unitary coupled cluster method. To understand the source of this residual error, we have reperformed the FNS-UCC3 and FNS-qUCCSD calculation in an uncontracted aug-cc-pVQZ basis set (dyall.acv4z basis set is used for Br and I). (See Table S1). It can be seen that dipole moment values slightly increase on going from TZ to QZ basis set, especially for HBr and HI, and their result shows excellent agreement with experimental results with an error of ~0.01 Debye. The FNS-UCC3 and FNS-qUCCSD give almost similar results, with the latter giving slightly superior performance in cases.

To understand the effect of relativity on UCC3 and qUCCSD dipole moments, we have reperformed the calculations using a non-relativistic Hamiltonian. From Table S1, it can be seen that the effect of relativity does not lead to any appreciable change for HF and HCl with a small deviation (~0.02 Debye) in both FNS-UCC3 and FNS-qUCCSD methods. However, the inclusion of the relativistic effect leads to a significant decrease in the dipole moment value as we go down the periodic table. The deviation between the relativistic and non-relativistic results is 0.05 Debye for HBr and 0.14 Debye for HI, both for UCC3 and qUCCSD methods, respectively. The inclusion of the relativistic effect improved the agreement with the experiment except in HF. The most prominent effect is observed for HI, where the error in non-relativistic FNO-qUCCSD with respect to the experiment is 0.13 Debye, which improves to 0.01 Debye on the inclusion of the relativistic effect.

### 3.3 Bond length

Bond length and harmonic vibrational frequencies in this paper have been calculated using the finite differentiation method. The fifth-degree polynomial is used, and the convergence of the DHF and coupled cluster energy is taken up to $10^{-12}$ a.u. Table 3 presents canonical fc-UCC3 and fc-qUCCSD bond lengths for HX (X=F, Cl, Br, I). Uncontracted aug-cc-pVTZ basis set (dyall.acv3z basis set for Br and I) has been used for the calculations. It can be seen that the results in both canonical UCC3 and qUCCSD methods show excellent agreement with the corresponding CCSD values. The agreement is better for the qUCCSD method, where the maximum deviation is within 0.0005 Å of CCSD values for HCl molecule. The UCC3 shows a slightly higher error with a maximum deviation of 0.0013 Å for HI molecule. The deviation between the relativistic unitary coupled cluster methods with its standard CCSD analogue is much smaller than the error observed in the computed value with respect to the experiment. The maximum deviation is observed for HI, where UCC3 and qUCCSD methods show errors of 0.0078 Å and 0.0064 Å, respectively, with respect to the experimental bond length (1.6092 Å for HI). The standard CCSD method shows a deviation of 0.0065 Å from the experiment. Figure 6 presents the convergence of the bond length of the HCl molecule with respect to the size of the virtual space for four component UCC3 and qUCCSD methods in canonical and FNS basis. It can be seen that, unlike the dipole moment, the error in bond length with a truncated virtual space is much lower in the FNS basis than that observed in the canonical basis. The FNS-UCC3 and FNS-qUCCSD results converge at ~70% and ~75% of the virtual space, whereas the canonical results converge at around ~80% of the virtual space. The inclusion of perturbative correction significantly improves the convergence, and the bond length in both FNS-UCC3 and FNS-qUCCSD converges at ~40% of the virtual space when perturbative corrections are considered. Figure 7 presents the convergence of the FNS-UCC3 and FNS-qUCCSD bond lengths with respect to the FNS truncation threshold in frozen core approximation. It can be seen that bond length converges at a $10^{-6}$ threshold without perturbative correction. The inclusion of perturbative correction improves the convergence, and the error becomes negligible even at the $10^{-5}$ threshold. Table 3 shows that the use of FNS approximation introduces negligible error in four component UCC3 and qUCCSD methods for HX (X=F, Cl, Br, I) molecules. The maximum error observed in both the methods is below 0.001 Å, and they get reduced below 0.0002 Å on the inclusion of perturbative correction. The use of a quadruple zeta quality basis set leads to a slight improvement of the results (see Table S2), with the maximum error observed (for HI) with respect to the experiment in FNS-UCC3 and FNS-qUCCSD methods is reduced by 0.007 Å and 0.0058 Å respectively, when perturbative correction is considered.

We have also calculated the bond length in the non-relativistic variant of the UCC3 and qUCCSD methods in the uncontracted aug-cc-pVQZ basis set (dyall.acv4z basis set for Br and I ). It can be seen that the inclusion of the relativistic effect leads to the contraction of bond length, as previously reported for standard CCSD by Visscher and co-workers[9], except in the case of HF, where the effect of the relativity seems to be negligible. The effect is generally very small even for HCl, and the decrease is 0.0001 Å due to the inclusion of the relativistic effect. The effect becomes appreciable for HBr, where the inclusion of the relativistic effect leads to a decrease of 0.002 Å in bond length in both FNS-UCC3 and FNS-qUCCSD methods. The HI shows an even larger reduction in bond length of 0.005 Å on the inclusion of the relativistic effect. One can see that the inclusion of the relativistic effect does not necessarily improve the agreement of the UCC3 and qUCCSD results with the experiment.

**3.4 Harmonic vibrational frequency**

Table 4 presents the harmonic vibrational frequencies for HX (X=F, Cl, Br, I) series in relativistic UCC3 and qUCCSD methods. Uncontracted aug-cc-pVTZ basis set (dyall.acv3z basis set for Br and I ) has been used for the calculations. The results are generally in good agreement with the corresponding standard relativistic CCSD method. The maximum deviation obtained from the corresponding CCSD is around 15 cm$^{-1}$ and 8 cm$^{-1}$ in UCC3 and qUCCSD methods, respectively. Both these deviations are much less than the errors with respect to the corresponding experimental values. The standard relativistic CCSD method shows a maximum deviation of 26

cm$^{-1}$ from the experiment. The maximum deviations obtained in UCC3 and qUCCSD with respect to the experiment are slightly higher at 40 cm$^{-1}$ and 31 cm$^{-1}$, for HBr and HF respectively.

Figure 8 presents the convergence of the harmonic vibration frequency for the HCl molecule with respect to the size of the virtual space for four-component UCC3 and qUCCSD methods in the canonical and FNS basis. It can be seen that, unlike the dipole moment, the error in the vibrational frequency with a truncated virtual space convergence is much faster in the FNS basis than that observed in the canonical basis. The FNS-UCC3 and FNS-qUCCSD methods converge at a virtual space size of 60% and 70%, respectively. The inclusion of perturbative correction significantly improves the convergence and the vibrational frequencies in both FNS-UCC3 and FNS-qUCCSD methods, which converge at around 40% of the virtual space. Figure 9 presents the convergence of the FNS-UCC3 and FNS-qUCCSD vibrational frequencies with respect to the FNS truncation threshold and the results converge at a threshold of 10$^{-6}$. The inclusion of perturbative correction results in quicker convergence of vibrational frequencies, and UCC3 and qUCCSD results both converge at the 10$^{-5}$ threshold.

Table 4 shows that the use of FNS approximation leads to negligible error in four component UCC3 and qUCCSD methods for HX (X=F, Cl, Br, I) molecules. The maximum errors observed for HBr in UCC3 and qUCCSD are 17 cm$^{-1}$ and 22 cm$^{-1}$ with respect to the corresponding untruncated canonical values, respectively. Upon the inclusion of perturbative correction, the errors are reduced to 10 cm$^{-1}$ and 3 cm$^{-1}$, respectively. The errors with respect to the experiment for both UCC3 and qUCCSD methods are higher than the FNS errors. The errors are comparable to those observed in the standard relativistic CCSD method. The increase in the basis set to uncontracted aug-cc-pVQZ (dyall.acv4z basis set for Br and I) in the FNS-qUCCSD method leads to a decrease in the harmonic vibrational frequency for all the molecules, except that of the HF molecule, with a reduction of errors with respect to the experimental results (see Table S3). The error in the case of HF increases on going to the aug-cc-pVQZ basis set. The change in the FNS-UCC3 results on going to QZ basis set shows a similar trend. We have also calculated the FNO-UCC3 and FNO-qUCCSD results in uncontracted aug-cc-pVQZ (dyall.acv4z basis set for Br and I) using a non-relativistic Hamiltonian. The inclusion of the relativistic effect leads to a decrease in harmonic vibrational frequency, similar to that reported by Visscher and coworkers[9]. The effect is more prominent as we go down the group in the periodic table.

**3.5 Computational efficiency**

To investigate the computational performance of the relativistic UCC3 and qUCCSD methods along with its FNS versions, we have calculated the ground state correlation energy of the HBr molecule. The uncontracted aug-cc-pVTZ basis set is used for H and dyall.acv3z basis set is used for Br. All the calculations were performed on a dedicated workstation having two Intel(R) Xeon(R) CPU E5-2667 v4 @3.20 GHz CPU and 512 GB of total RAM. The individual computational timings are provided in Table S4. The correlation part in standard relativistic CCSD method takes 13 hours 29 minutes, and 24 seconds, out of which 4 hours 11 minutes and 24 seconds are spent in integral transformation and 8 hours 51 minutes and 36 seconds in coupled-cluster iterations.

The UCC3 and qUCCSD methods take almost similar time as the standard relativistic CCSD calculation for the integral transformation. The coupled cluster iterations take 4 hours, 10 minutes, 12 seconds, and 15 hours, 27 minutes, 36 seconds for UCC3 and qUCCSD methods, respectively, which is in line with the cost analysis reported in Section 2.1. The use of FNS approximation reduces the number of virtual orbitals from 362 to 178 and also the computational cost of both integral transformation and coupled cluster iterations. The timing for FNS generation and integral transformation for both UCC3 and qUCCSD is 1 hour and 36 minutes, which is around three times less than the corresponding canonical integral transformation time. The coupled cluster iteration time was reduced to 30 minutes 12 seconds and 1 hour 2 minutes 31 seconds for UCC3 and qUCCSD, respectively, using the FNS approximation. The total computational time for correlation calculation in the UCC3 and qUCCSD methods undergoes 4 and 7.5 fold reductions using FNS approximation (See Figure (10)).

## 4. Conclusions:

We present the theory, implementation, and benchmarking of the four-component relativistic unitary coupled cluster method based on commutator-based non-perturbative approximation for molecules. The quadratic unitary coupled cluster (qUCCSD) and its perturbative approximation (UCC3) have been reported. Both methods show their performance comparable to that of the standard coupled cluster method for dipole moment, bond length, and harmonic vibrational frequency calculations. The qUCCSD method shows a slightly better agreement with the CCSD results than the UCC3, presumably due to a more complete treatment of the singles containing quadratic terms. In terms of computational cost, the relativistic UCC3 iterations take less time than the corresponding standard relativistic CCSD method. The computation cost of qUCCSD iterations is around twice that of the standard relativistic CCSD method. The transformation of two-electron integrals from atomic to molecular spinor basis accounts for a large chunk of the total computation time. The use of FNS approximation can significantly reduce the computational cost of both integral transformation and coupled cluster iteration steps, with a higher speed-up observed for the qUCCSD method. An FNS threshold of $10^{-5}$ gives an economic compromise between accuracy and efficiency. The use of perturbative correction is essential for getting smooth convergence of the properties in both UCC3 and qUCCSD methods. The calculated results show strong dependence upon the used basis set. The qUCCSD method shows better agreement with experimental results than the UCC3 method for dipole moment, bond length, and IR frequency. However, none of the methods gives quantitive agreement with the experiment for bond length and frequency, presumably due to missing higher-order correlation correction. The inclusion of the relativistic effect leads to a contraction in the bond length and a reduction in the dipole moment and harmonic vibrational frequency values in unitary coupled cluster methods. The effect is more prominent as we go down the periodic table. Extending the relativistic unitary coupled cluster method to perturbative triples correction, excited states, and analytic calculation of properties will be necessary to use them for practical applications. Work is in progress towards that direction.

**Supplementary Material:** The explicit programable expressions for UCC3 and qUCCSD methods, comparison of non-relativistic and relativistic results for dipole moment, bond length, and harmonic vibrational frequency, and details of the timing comparison have been provided in the supporting information.

**Acknowledgment:** The authors acknowledge the support from the IIT Bombay, DST-SERB CRG (Project No. CRG/2022/005672) and MATRICS (Project No. MTR/2021/000420) projects, DST-Inspire Faculty Fellowship (Project No. DST/INSPIRE/04/2017/001730), CSIR-India (Project No. 01(3035)/21/EMR-II), ISRO (Project No. R.D./0122-ISROC00-004), and IIT Bombay super computational facility and C-DAC supercomputing resources (PARAM Smriti and PARAM Brahma) for computational time. KM acknowledges CSIR-HRDG for senior research fellowship.

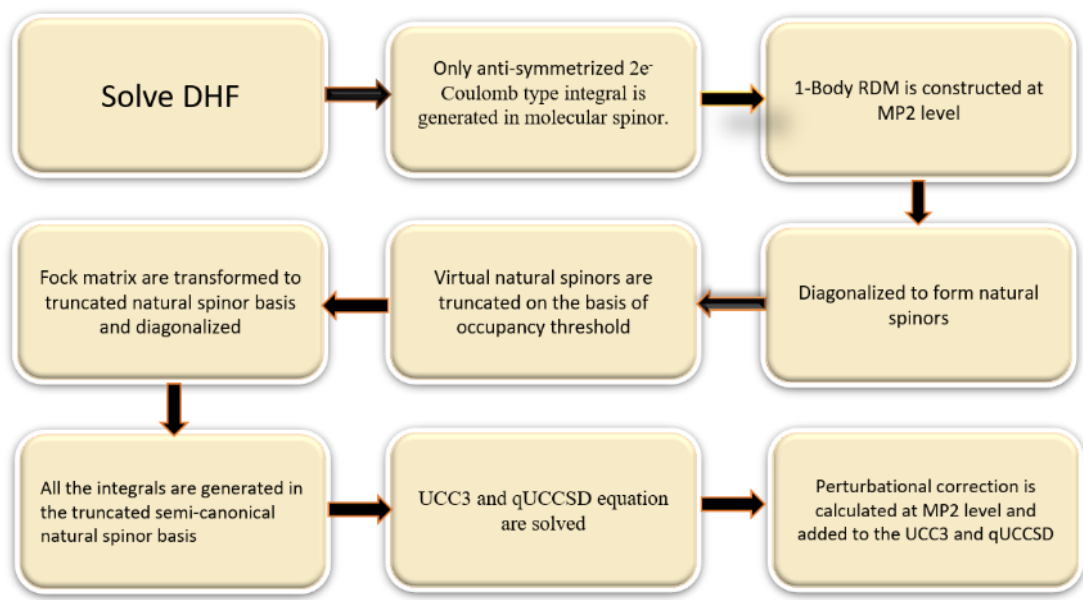

*Figure 1: A schematic representation of the algorithm for FNS-UCC3 and FNS-qUCCSD.*

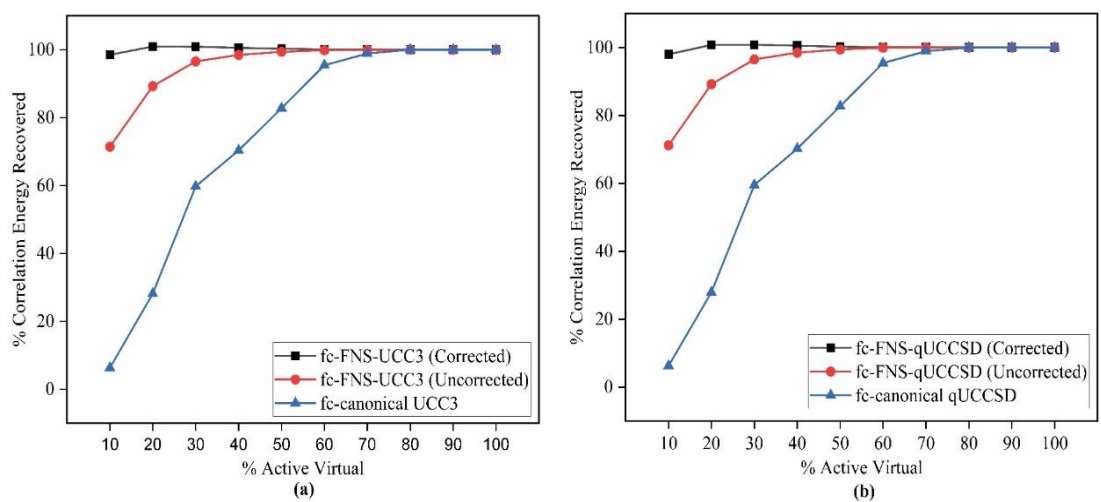

*Figure 2: The convergence plot of correlation energy with respect to the size of the virtual space in canonical and FNS basis with frozen core (fc) approximation in the four-component relativistic (a) UCC3 and (b) qUCCSD methods for the HCl molecule using uncontracted aug-cc-pVTZ basis set.*

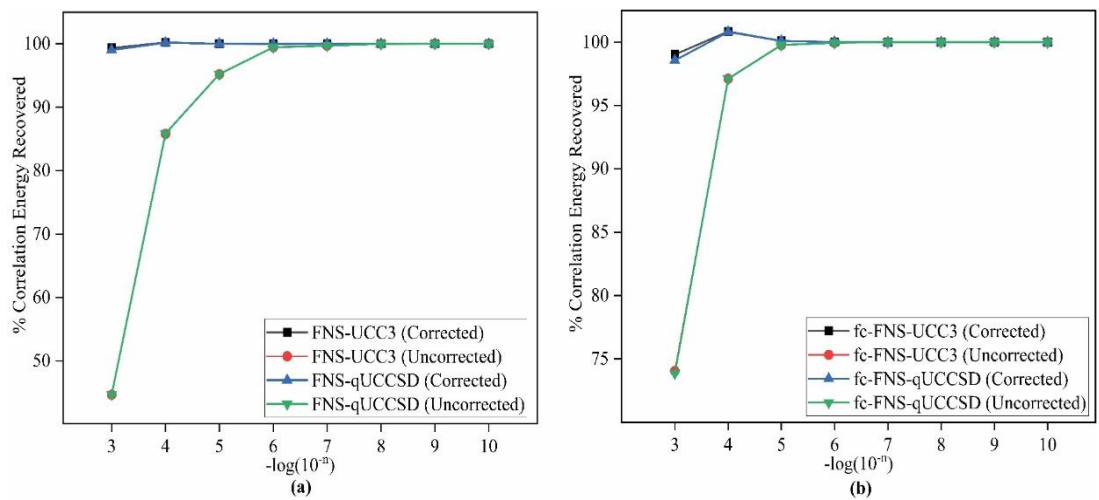

*Figure 3: The convergence plot of correlation energy calculated by FNS-UCC3 and FNS-qUCCSD methods (a) without and (b) with frozen core (fc) approximation with respect to the FNS truncation threshold ($10^{-n}$) for HCl molecule using uncontracted aug-cc-pVTZ basis set.*

*Table 1: The percentage of the correlation energy recovered in FNS-UCC3 and FNS-qUCCSD methods with an FNS truncation threshold of $10^{-5}$ with frozen core (fc) approximation using uncontracted aug-cc-pVTZ basis set (dyall.acv3z has been used for Br and I).*

| Molecule | Total virtual spinors | Active virtual spinors | % Error in correlation energy recovered w.r.t. canonical energy | | | |
|:---:|:---:|:---:|:---:|:---:|:---:|:---:|
| | | | FNS-UCC3 (Uncorrected) | FNS-UCC3 (Corrected) | FNS-qUCCSD (Uncorrected) | FNS-qUCCSD (Corrected) |
| HF | 156 | 94 | -0.41 | 0.08 | -0.39 | 0.10 |
| HCl | 182 | 104 | -0.22 | 0.10 | -0.22 | 0.09 |
| HBr | 362 | 178 | -0.96 | 0.15 | -0.95 | 0.14 |
| HI | 434 | 174 | -0.67 | 0.14 | -0.67 | 0.13 |

*Table 2: The dipole moment (Debye) calculated using fc-UCC3 and fc-qUCCSD methods along with its FNS versions (threshold $10^{-5}$) in uncontracted aug-cc-pVTZ basis set (dyall.acv3z is used for Br and I).*

| Molecules | Canonical CCSD[a] | Canonical UCC3 | Error in dipole moment w.r.t. canonical UCC3 | | Canonical qUCCSD | Error in dipole moment w.r.t. canonical qUCCSD | | Exp[54] |
|---|---|---|---|---|---|---|---|---|
| | | | FNS-UCC3 (Uncorrected) | FNS-UCC3 (Corrected) | | FNS-qUCCSD (Uncorrected) | FNS-qUCCSD (Corrected) | |
| HF | 1.8055 | 1.806 | -0.008 | -0.001 | 1.804 | 0.007 | 0.002 | 1.82 |
| HCl | 1.0826 | 1.086 | 0.005 | -0.002 | 1.084 | 0.006 | -0.002 | 1.08 |
| HBr | 0.7937 | 0.797 | 0.017 | -0.002 | 0.795 | 0.017 | -0.002 | 0.82 |
| HI | 0.4003 | 0.402 | 0.000 | -0.002 | 0.401 | 0.000 | -0.002 | 0.44 |

[a]Taken from reference[40]

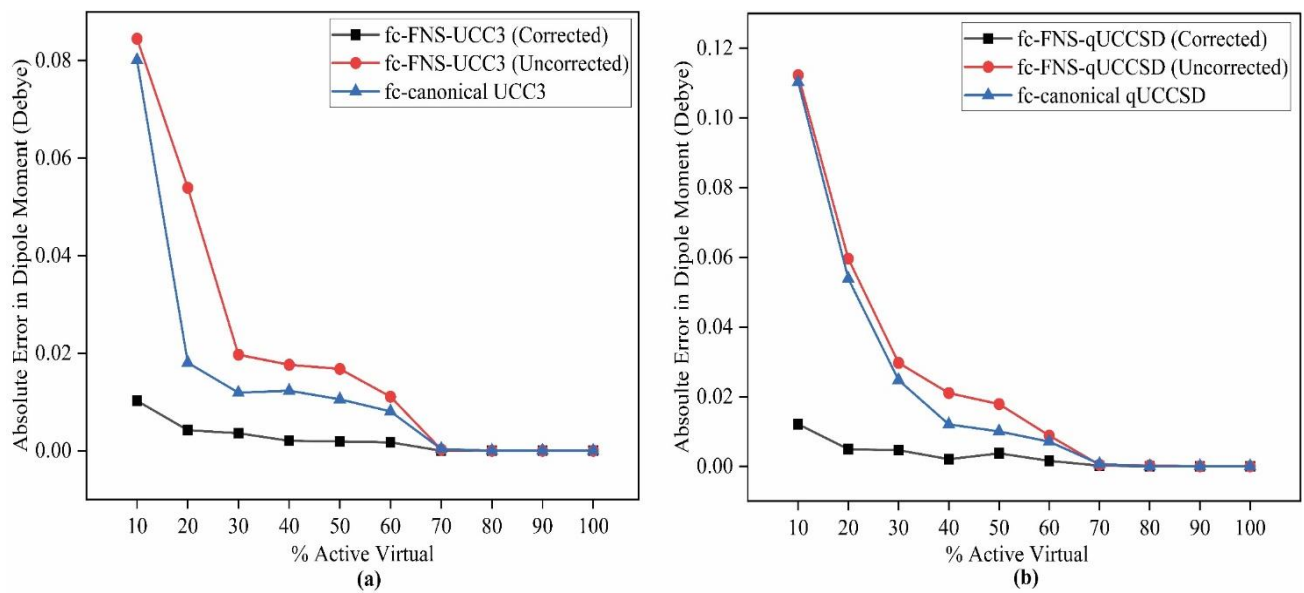

*Figure 4: The convergence plot of the dipole moment calculated by the relativistic four-component (a) UCC3 and (b) qUCCSD methods with respect to the size of the virtual space in the canonical and the FNS basis with frozen core (fc) approximation for HCl molecule in uncontracted aug-cc-pVTZ basis set.*

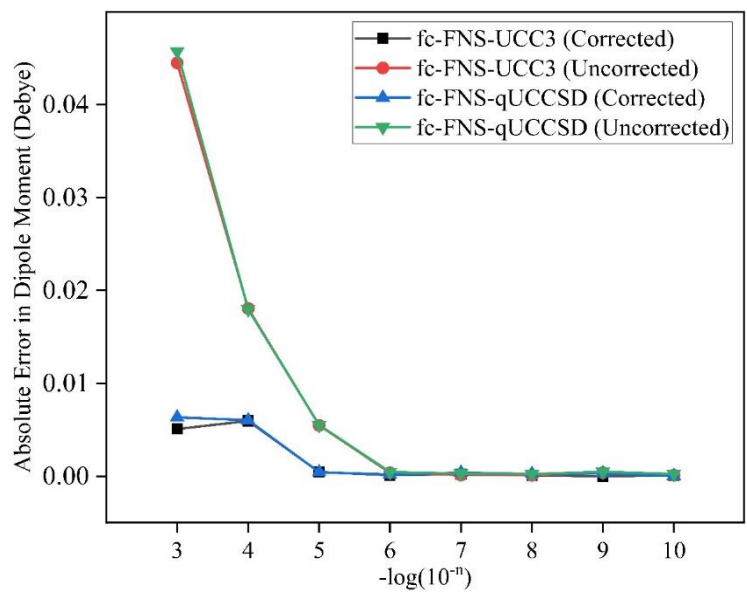

*Figure 5: The convergence of the dipole moment calculated using FNS-UCC3 and FNS-qUCCSD methods with respect to the FNS truncation threshold ($10^{-n}$) for HCl molecule in uncontracted aug-cc-pVTZ with frozen core (fc) approximation.*

*Table 3: The bond length (Å) calculated using fc-UCC3 and fc-qUCCSD methods along with its FNS versions (threshold $10^{-5}$) in uncontracted aug-cc-pVTZ basis set (dyall.acv3z is used for Br and I).*

| Molecules | Canonical CCSD[a] | Canonical UCC3 | Error in bond length w.r.t. canonical UCC3 | | Canonical qUCCSD | Error in bond length w.r.t. canonical qUCCSD | | Exp[55] |
|---|---|---|---|---|---|---|---|---|
| | | | FNS-UCC3 (Uncorrected) | FNS-UCC3 (Corrected) | | FNS-qUCCSD (Uncorrected) | FNS-qUCCSD (Corrected) | |
| HF | 0.9178 | 0.9174 | -0.0005 | 0.0001 | 0.9176 | -0.0005 | 0.0001 | 0.9168 |
| HCl | 1.2753 | 1.2751 | -0.0002 | 0.0001 | 1.2758 | -0.0002 | 0.0001 | 1.2746 |
| HBr | 1.4089 | 1.4078 | -0.0004 | 0.0002 | 1.4089 | -0.0007 | 0.0001 | 1.4144 |
| HI | 1.6027 | 1.6014 | -0.0004 | 0.0001 | 1.6028 | -0.0005 | 0.0001 | 1.6092 |

[a]Taken from reference[40]

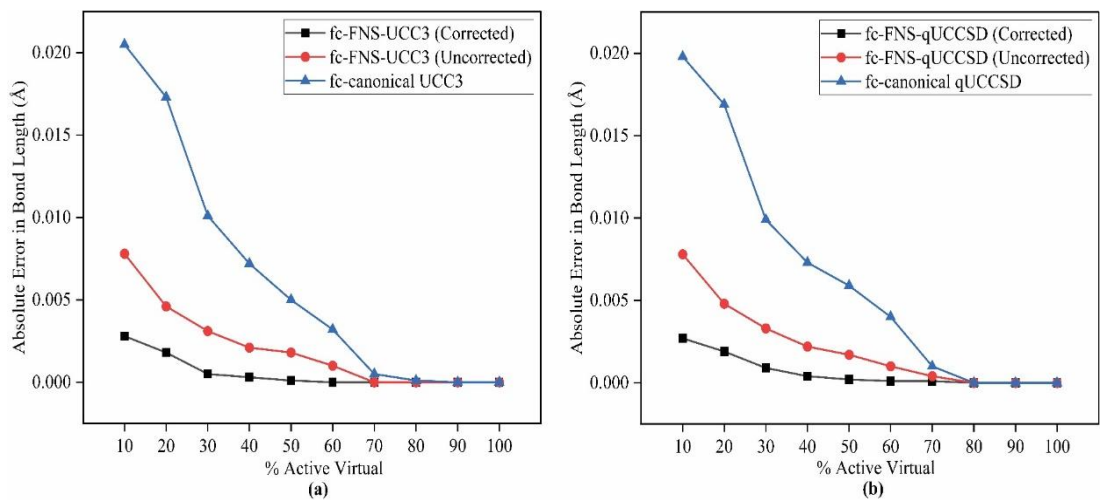

*Figure 6: The convergence of bond length values of HCl with respect to the size of the virtual space in canonical and FNS basis in relativistic four-component (a) UCC3 and (b) qUCCSD with frozen core (fc) approximation in uncontracted aug-cc-pVTZ basis set.*

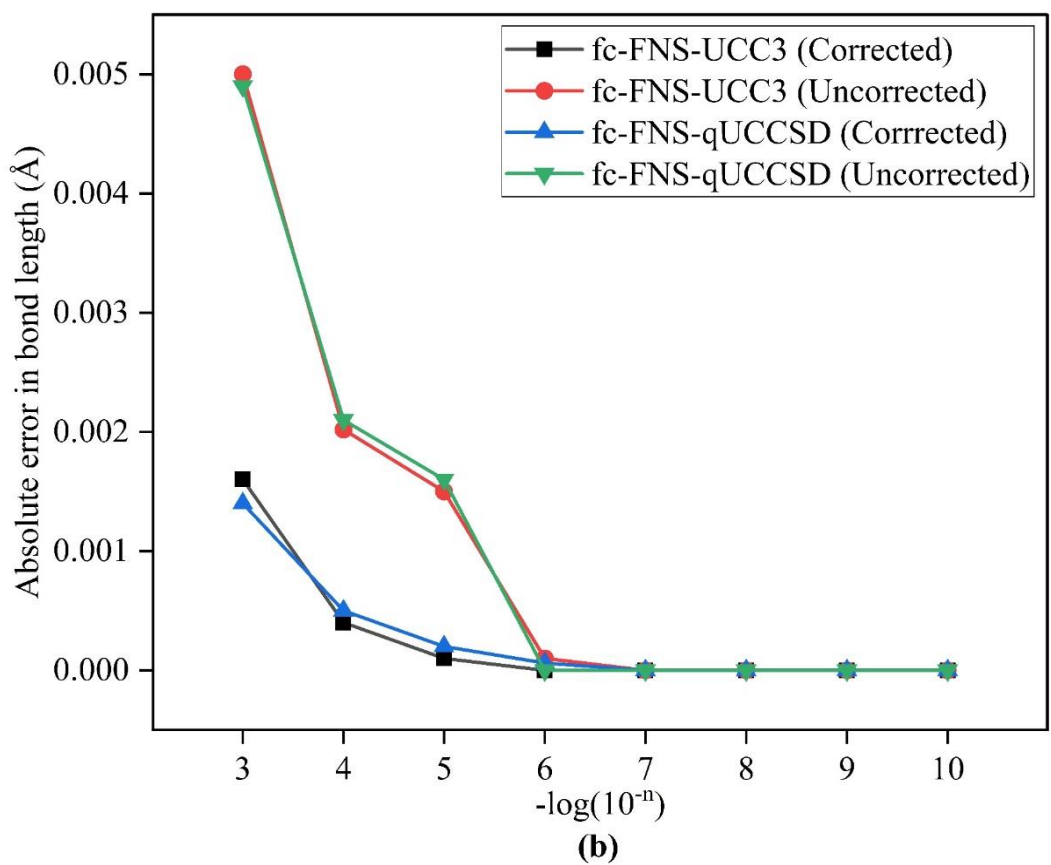

*Figure 7: The convergence of bond length calculated using the FNS-UCC3 and FNS-qUCCSD methods with respect to the FNS truncation threshold ($10^{-n}$) for HCl molecule with frozen core (fc) approximation.*

*Table 4: The harmonic vibrational frequency (cm$^{-1}$) calculated using fc-UCC3 and fc-qUCCSD methods along with its FNS versions (threshold 10$^{-5}$) in uncontracted aug-cc-pVTZ basis set (dyall.acv3z is used for Br and I).*

| Molecules | Canonical CCSD[a] | Canonical UCC3 | Error in frequency w.r.t. canonical UCC3 | | Canonical qUCCSD | Error in frequency w.r.t. canonical qUCCSD | | Exp[55] |
|---|---|---|---|---|---|---|---|---|
| | | | FNS-UCC3 (Uncorrected) | FNS-UCC3 (Corrected) | | FNS-qUCCSD (Uncorrected) | FNS-qUCCSD (Corrected) | |
| HF | 4164.15 | 4174.48 | 9.65 | -1.38 | 4169.19 | -9.9 | -1.62 | 4138.32 |
| HCl | 3015.13 | 3022.56 | 2.26 | -0.58 | 3011.97 | 1.25 | 0.81 | 2990.94 |
| HBr | 2673.10 | 2688.56 | 17.40 | -9.62 | 2676.41 | 22.7 | -2.91 | 2648.97 |
| HI | 2333.70 | 2332.45 | 27.60 | -1.90 | 2325.80 | 12.40 | -4.5 | 2309.01 |

[a]Taken from reference[40]

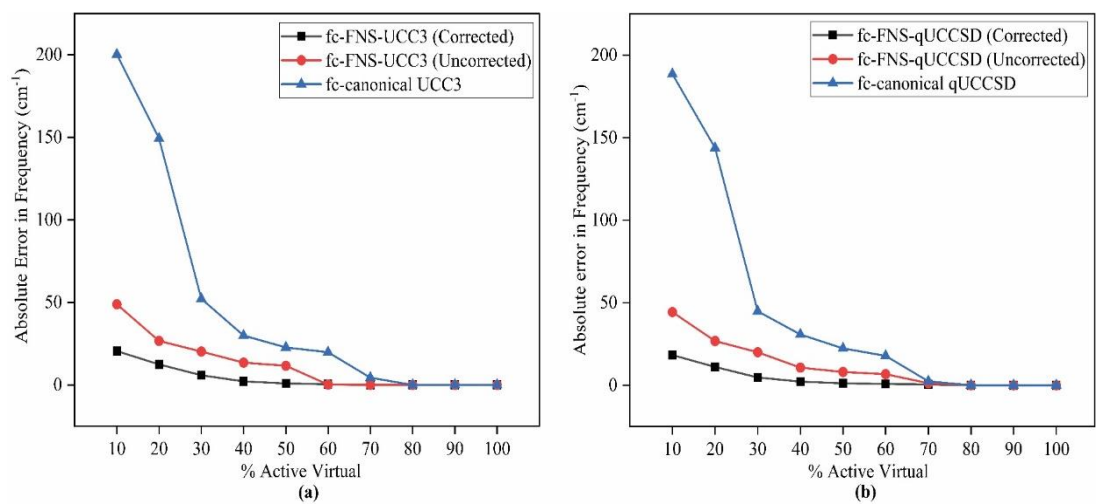

*Figure 8: The convergence of harmonic vibrational frequency values of HCl with respect to the size of the virtual space in canonical and FNS basis in relativistic four-component (a) UCC3 and (b) qUCCSD with frozen core (fc) approximation in uncontracted aug-cc-pVTZ basis set.*

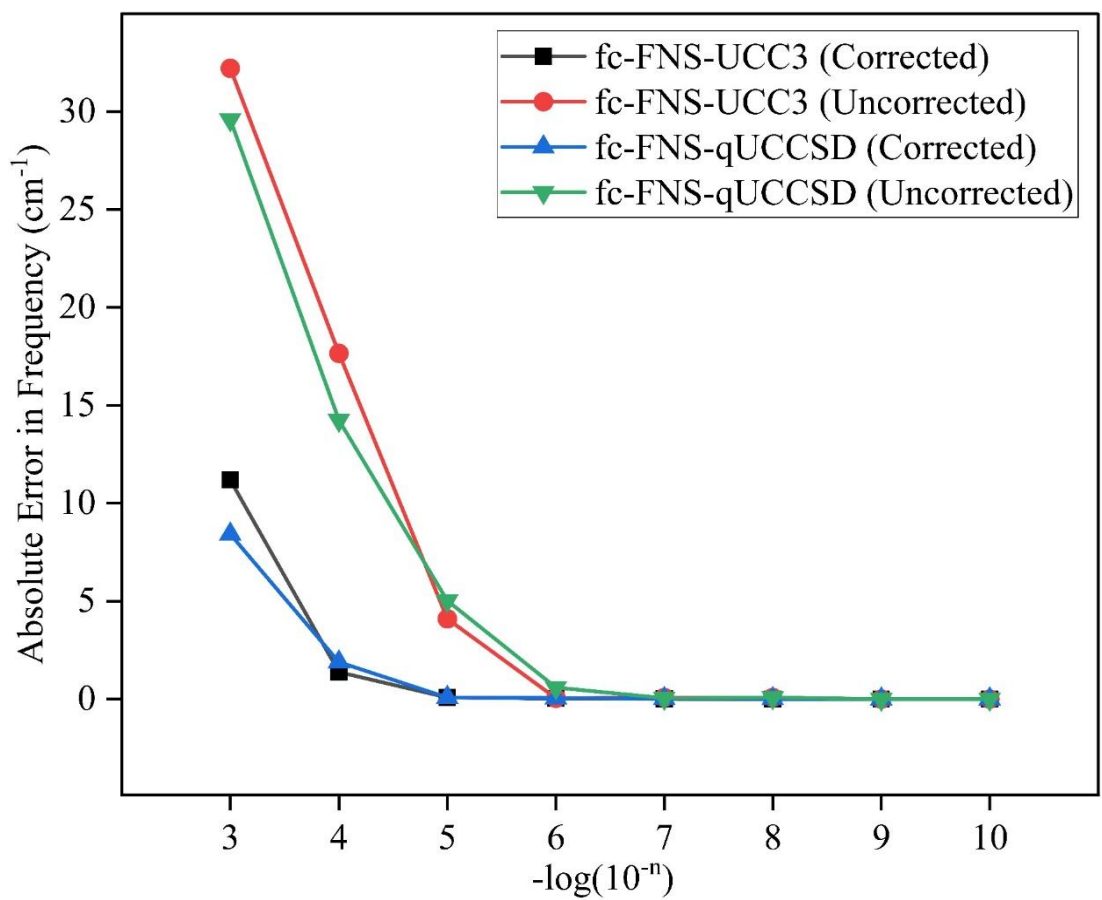

*Figure 9: The convergence of the harmonic vibrational frequency calculated using the FNS-UCC3 and FNS-qUCCSD methods with respect to the FNS truncation threshold ($10^{-n}$) for HCl molecule with frozen core (fc) approximation in uncontracted aug-cc-pVTZ basis set.*

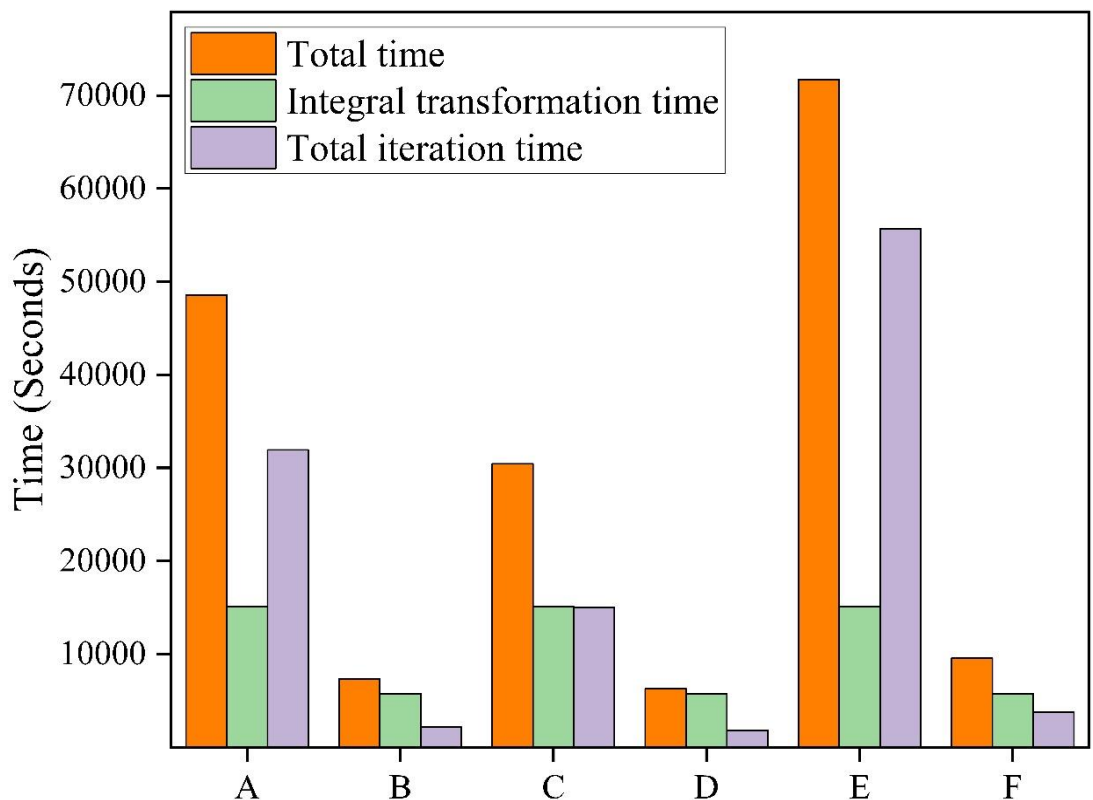

*Figure 10: Comparison of time taken by different methods for the calculation of correlation energy in fc-canonical and fc-FNS based implementations for HBr molecule. The uncontracted aug-cc-pVTZ basis set is used for H and dyall.acv3z basis set is used for Br. The processes A, B, C, D, E and F are as follows: (A) CCSD; (B) FNS-CCSD; (C) UCC3; (D) FNS-UCC3; (E) qUCCSD and (F) FNS-qUCCSD.*